\documentclass[sigconf]{acmart}

\AtBeginDocument{%
  }

\acmConference[AI-ML Systems]{$4^{th}$ International Conference on AI-ML Systems}{October 08--11, 2024}{Louisiana State University, Baton Rouge, USA}

\usepackage{amsmath}
\usepackage{amsthm}
\usepackage{booktabs}
\usepackage{enumitem}
\usepackage{graphicx}
\usepackage{color}
\usepackage{caption}
\usepackage{subcaption}
\usepackage{multirow}
\usepackage{multicol}
\usepackage{ragged2e}

\acmYear{2024}
\acmConference[AIMLSystems 2024]{4th International Conference on AI-ML Systems}{October 08--11, 2024}{Baton Rouge, LA, USA}
\acmBooktitle{4th International Conference on AI-ML Systems (AIMLSystems 2024), October 08--11, 2024, Baton Rouge, LA, USA}
\acmPrice{}
\acmDOI{10.1145/3703412.3703420}
\acmISBN{979-8-4007-1161-9/24/10}

\setcopyright{none}
\thanks{This is the author’s version of the work. The definitive Version of Record is published in the ACM Digital Library at \url{https://doi.org/10.1145/3703412.3703420}.}

\begin{document}


\title[Quantifying Cryptocurrency Unpredictability]{Quantifying Cryptocurrency Unpredictability: \\A Comprehensive Study of Complexity and Forecasting}

\author{Francesco Puoti}
\email{francesco.puoti@polimi.it}
\orcid{0009-0006-4661-3613}
\affiliation{%
  \institution{Politecnico di Milano}
  \department{Department of Electronics Information and Bioengineering}
  \city{Milano}
  \country{Italy}
}

\author{Fabrizio Pittorino}
\email{fabrizio.pittorino@polimi.it}
\orcid{0000-0002-1919-6141}
\affiliation{%
  \institution{Politecnico di Milano}
  \department{Department of Electronics Information and Bioengineering}
  \city{Milano}
  \country{Italy}
}

\author{Manuel Roveri}
\email{manuel.roveri@polimi.it}
\orcid{0000-0001-7828-7687}
\affiliation{%
  \institution{Politecnico di Milano}
  \department{Department of Electronics Information and Bioengineering}
  \city{Milano}
  \country{Italy}
}

\renewcommand{\shortauthors}{Puoti, Pittorino, Roveri}

\begin{abstract}
This paper offers a thorough examination of the univariate predictability in cryptocurrency time-series. By exploiting a combination of complexity measure and model predictions we explore the cryptocurrencies time-series forecasting task focusing on the exchange rate in USD of Litecoin, Binance Coin, Bitcoin, Ethereum, and XRP. On one hand, to assess the complexity and the randomness of these time-series, a comparative analysis has been performed using Brownian and colored noises as a benchmark. The results obtained from the Complexity-Entropy causality plane and power density spectrum analysis reveal that cryptocurrency time-series exhibit characteristics closely resembling those of Brownian noise when analyzed in a univariate context. On the other hand, the application of a wide range of statistical, machine and deep learning models for time-series forecasting demonstrates the low predictability of cryptocurrencies. Notably, our analysis reveals that simpler models such as Naive models consistently outperform the more complex machine and deep learning ones in terms of forecasting accuracy across different forecast horizons and time windows. The combined study of complexity and forecasting accuracies highlights the difficulty of predicting the cryptocurrency market. These findings provide valuable insights into the inherent characteristics of the cryptocurrency data and highlight the need to reassess the challenges associated with predicting cryptocurrency's price movements.
\end{abstract}

\ccsdesc[500]{Applied computing~Mathematics and statistics}
\ccsdesc[500]{Computing methodologies~Neural networks}
\ccsdesc[500]{Applied computing~Economics}
\ccsdesc[500]{Mathematics of computing~Time series analysis}

\keywords{Time series Forecasting, Time series Complexity, Time series Entropy, Cryptocurrencies, Preemptive Analysis, Machine Learning, Deep Learning, Naive Models}

\begin{teaserfigure}
  \centering
  \includegraphics[width=0.8\textwidth, height=0.2\paperheight]{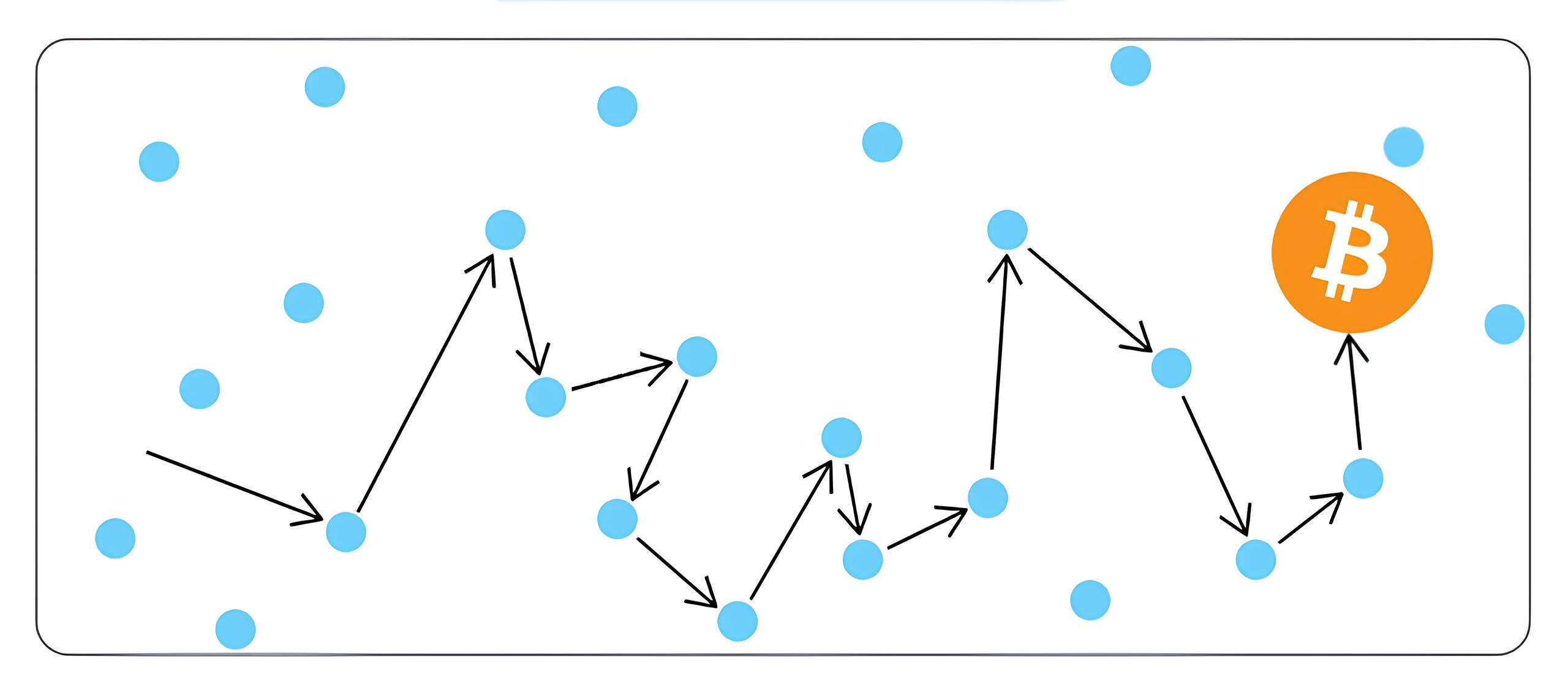}
  \caption{Cryptocurrency as Brownian Motion}
  \Description{The image illustrates the concept of cryptocurrency price movements through the analogy of Brownian motion. It features a rectangular frame filled with randomly scattered light blue dots. A black zigzag path with arrows connects several of these dots, symbolizing the unpredictable fluctuations in cryptocurrency prices over time. At the end of this path, an orange circle with the Bitcoin symbol highlights the resemblance of cryptocurrency time-series to Brownian motion.}
  \label{fig:teaser}
\end{teaserfigure}


\maketitle

\section{Introduction}
Cryptocurrencies have emerged as a significant asset in the global financial system, offering a new paradigm for digital transactions~\citep{nakamoto2008bitcoin}. Their decentralized nature, coupled with their potential for high returns, has attracted the attention of investors, researchers, and financial institutions alike~\citep{BAUR2018177}. 
The cryptocurrency market is influenced by several key factors. Market sentiment and speculation play a significant role, with prices being highly sensitive to social media, news events, celebrity endorsements, and public perceptions. These speculative behaviors cause what are perceived as rapid and unpredictable price changes, making it difficult to distinguish between genuine value shifts and short-term market reactions~\citep{cryptosurvey}.
Additionally, the evolving and inconsistent regulatory environment is another influential factor. Announcements of new regulations, enforcement actions, or changes in legal status can lead to sudden and volatile price movements, creating an unstable investment landscape. 
This volatile and complex behavior of cryptocurrencies presents unique challenges for traditional financial analysis and forecasting methods~\citep{Akyildirim2021}.

This study analyzes on an extended time period five prominent cryptocurrencies' exchange rates in USD: Litecoin (LTC-USD), Binance Coin (BNB-USD), Bitcoin (BTC-USD), XRP (XRP-USD), and Ethereum (ETH-USD). We aim to provide a comprehensive understanding of cryptocurrency dynamical features and put under scrutiny their predictability in a univariate context.
To this aim, this study leverages complexity measures such as the Permutation Entropy~\citep{pe} and the CH-plane~\citep{statisticalcomplexity}, and statistical, machine and deep learning models for time-series analysis, ranging from simple Naive models and ARIMA~\citep{Hyndman2021} up to more complex models such as XGBModels~\citep{Chen2016} and NBEATS~\citep{Oreshkin2019}. 
This approach combines advanced statistical complexity methods with state-of-the-art machine learning techniques to capture the intricate patterns and potential predictability in cryptocurrency markets~\citep{LAHMIRI2022118349}.
Our methodology includes the analysis of cryptocurrencies through the Complexity-Entropy causality plane (CH-plane)~\citep{statisticalcomplexity} and power density spectrum. These tools allow us to draw parallels with Brownian motion, a well-known stochastic process that describes the random motion of particles suspended in a fluid.
Interestingly, Brownian motion is often used as a benchmark for random behavior in financial time-series and has a long tradition in the modelling of the stock market~\citep{FaridaAgustini_2018}. By comparing the characteristics of cryptocurrency time-series to Brownian motion, this study aims to gain insights into the efficiency and randomness of these markets, ultimately addressing the question of whether cryptocurrencies exhibit predictable patterns or are essentially pure noise.
Furthermore, we employ a range of forecasting models, from simple statistical approaches to sophisticated machine and deep learning algorithms, to assess the predictability of cryptocurrency price movements~\citep{bocconi_paper}. This comparative analysis not only evaluates the performance of different forecasting techniques but also provides insights into the inherent predictability of cryptocurrency markets.
Our research emphasizes the potential effectiveness of simpler forecasting methods under certain conditions, challenging the hypothesis that more sophisticated models always yield better results in the context of cryptocurrency time-series forecasting~\citep{ZHANG2024108509}.
By combining complexity analysis with predictive modeling, this study aims to provide a comprehensive understanding of cryptocurrency dynamical behavior and predictability.

\section{Related literature}

Permutation entropy and the Complexity-Entropy (CH) plane have been widely used to characterize the complexity and predictability of time-series data. The concept of permutation entropy as a measure of the degree of randomness in a time-series was introduced in~\citep{pe}. This method has gained popularity due to its simplicity and effectiveness in quantifying the temporal structure of complex systems. The CH-plane, proposed in~\citep{statisticalcomplexity}, plots permutation entropy against the statistical Jensen-Shannon complexity measure $C_{JS}[P]$,  is a functional of the probability distribution~$P$ associated with the time series~\citep{pe}, allowing for the classification of different dynamical regimes. This approach has been successfully applied to various fields, including financial markets~\citep{ZUNINO20101891}.

The relationship between the CH-plane and the predictability of cryptocurrency time-series has been studied in several works. Notably, in~\citep{cryptocurrencydisorder} the authors analyzed the entropy and statistical complexity of Bitcoin and Ethereum time-series, suggesting that the price dynamics are largely driven by noise. This finding aligns with the efficient market hypothesis, which posits that asset prices fully reflect all available information, making them inherently unpredictable~\citep{TITAN2015442}.
Similarly, ~\citep{cryptocurrencydisorder} applied the CH-plane methodology to map cryptocurrencies and found that their behavior varies widely within the plane, with price dynamics ranging from stochastic to more structured. This work observed that cryptocurrencies with high market capitalization tend to be more complex and less entropic than those with very low market capitalization, suggesting that major cryptocurrencies are less market efficient.
While these studies have employed complexity measures and the CH-plane to assess the nature of cryptocurrency time-series, they are based on much shorter time windows and have primarily focused on characterizing the dynamics rather than directly evaluating predictability.

Despite the challenges posed by the noisy nature of cryptocurrency time-series, some studies have reported successful predictions using various machine learning (ML) and deep learning (DL) models. However, it is crucial to note that most of these models were not compared with naive forecasting methods, or they introduced past and future covariates in the forecasting task, potentially inflating their perceived effectiveness.
For example~\citep{Bouteska2024} employed a range of ML models to predict cryptocurrency returns, reporting promising results. However, the study did not include a comparison with naive forecasting methods. The study~\citep{catania2019forecasting} used GARCH-type models to forecast cryptocurrency volatility, incorporating exogenous variables, which may have contributed to the model's performance.
In~\citep{LahmiriChaos} the authors applied a DL-based approach for cryptocurrency price prediction, reporting high accuracy. However, this study did not include a comparison with simpler forecasting methods. Similarly,~\citep{mcnally2018predicting} used Recurrent Neural Networks and Long Short-Term Memory networks for Bitcoin price prediction, but also lacked comparisons with baseline models.
The authors of~\citep{extraregr} integrated additional market indicators and sentiment analysis in their forecasting models, potentially improving predictions but deviating from a purely univariate approach.
More recently,~\citep{LAHMIRI2022118349} investigated the predictability of cryptocurrency trading volume using support vector regression (SVR) with different kernels. They found that SVR with radial basis function kernel outperformed other models for next-day trading volume prediction, while SVR with polynomial kernel was superior for next-week predictions. 
These studies highlight the importance of a careful model evaluation and comparison with simple benchmarks to assess the true predictive power of complex models in cryptocurrency forecasting. Our study aims to address this gap by providing a comprehensive comparison of various forecasting methods, including naive models, in a purely univariate context.

\section{Methodology}

\subsection{Data Collection}

The daily pricing data of the cryptocurrencies used in this study are collected from the Yahoo Finance database. 
Each cryptocurrency time-series $y$ ranges from 2020-07-03 to 2023-12-21, and no data preprocessing was applied. The data was split into two subsets, i.e., a training series $y_{train}$ and a test series $y_{target}$. The split date index is 2023-07-04, meaning that the last 180 data points represent~$y_{target}$.

To comprehensively evaluate the forecasting models (statistical, ML, and DL) and conduct the complexity analysis, we adopted a multiple timescale approach. This allows for a more in-depth understanding of cryptocurrency dynamics across various time horizons. Three different training time-window lengths ($t_w$) were applied on $y_{train}$, i.e., (i) $t_w$~=~3 years, (ii) $t_w$~=~1 year and (iii) $t_w$~=~6 months. Using these three different $t_w$ values allows us to identify whether there are long-term, mid-term, or short-term patterns, respectively, within the cryptocurrency time-series data. The~$y_{target}$ is the same for each considered time window $t_w$.

\begin{figure}
    \RaggedRight    
    \includegraphics[width=0.48\textwidth]{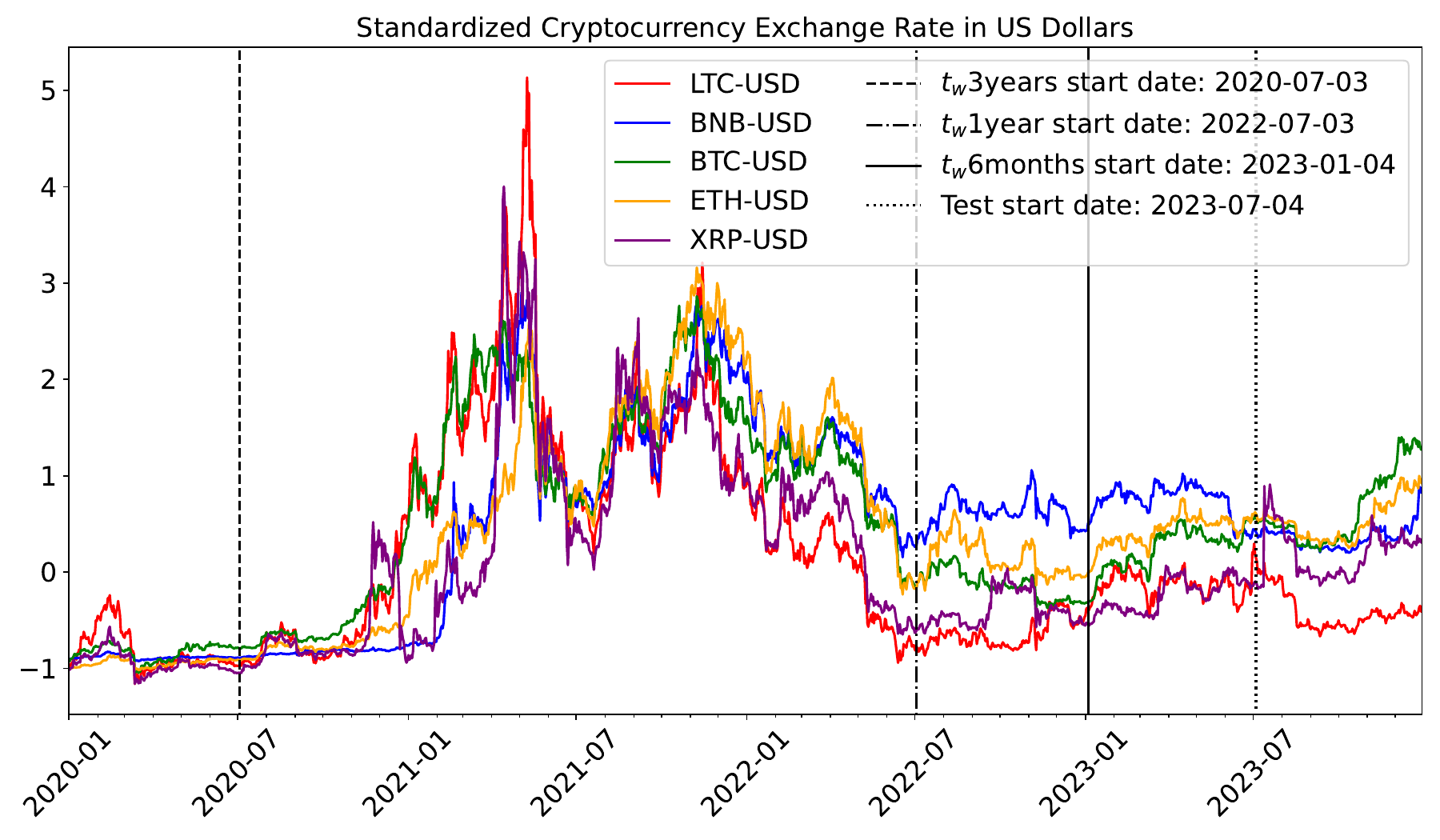}
    \caption{Overview of the Cryptocurrency time-series analyzed in this paper. The different starting dates relative to the three chosen time windows $t_w$ are shown. The cryptocurrency time-series have been standardized for clarity of the plot.}
    \label{fig:cryptoUSD}
\end{figure}

A visual summary of the cryptocurrency univariate time-series and scenarios considered in this paper is shown in Fig.~\ref{fig:cryptoUSD}. The considered scenarios are:
\begin{enumerate}
    \item $t_w$ = 3 years,\\${y_{train} = \{y(t) \, \forall \, \text{2020-07-03} \leq t \leq  \text{2023-07-04}\}}$
    \item $t_w$ = 1 year,\\${y_{train} = \{y(t) \, \forall \, \text{2022-07-03} \leq t \leq  \text{2023-07-04}\}}$
    \item $t_w$ = 6 months,\\${y_{train} = \{y(t) \, \forall \,  \text{2023-01-04} \leq t \leq  \text{2023-07-04}\}}$
\end{enumerate}

\subsection{Complexity Measures}
In order to evaluate the statistical complexity and the level of disorder and unpredictability of each cryptocurrency this study relies on the Bandt and Pompe \textit{permutation entropy}~\citep{pe} (PE) and the \textit{intensive statistical complexity measure} $C_{JS}[P]$ proposed in~\citep{statisticalcomplexity}, situating each cryptocurrency in the \textit{Complexity-Entropy} (CH) plane~\citep{statisticalcomplexity}.  

The permutation entropy quantifies the degree of randomness inherent in a process; the lower the entropy, the higher the predictability of the process. Focusing on the relative ordering of the time-series values, the permutation entropy takes into account the temporal causality within the series. For a time-series of length $n$, ordinal patterns of user-defined positive integer size $d : n\ge5d!$~\citep{cryptocurrencydisorder} are created. Then, the relative frequencies of these patterns are calculated to form a probability distribution. The permutation entropy is defined as the Shannon entropy of this distribution. It ranges from 0, which represents complete predictability, to $log(d!)$, which represents maximum randomness. 
In this study, we set $d=5$ for $t_w=$ 3 years, and $d=4$ for $t_w=$ 1 year and $t_w=$ 6 months. 

Furthermore, we employ the Jensen–Shannon complexity measure $C_{JS}[P]$ to quantify the complexity of the underlying probability distribution $P$ of the time-series. This measure integrates the concepts of permutation entropy and Jensen–Shannon divergence~($D_{JS}$), enabling a comprehensive capture of both the uncertainty and the structure of the distribution. The calculation of $C_{JS}[P]$ involves a two-step process: first, determining the $D_{JS}$ between the probability distribution $P$ and a uniform distribution, and then multiplying this divergence by the entropy of $P$. This approach provides valuable insights into the system's complexity by simultaneously accounting for two critical aspects: \textit{diversity}, which represents the amount of uncertainty or unpredictability in the distribution, and \textit{distinctiveness}, which quantifies the degree to which privileged fluctuations exist among those accessible to the system~\citep{cryptocurrencydisorder}.

By calculating both quantities, valuable insight can be gained regarding the distribution and the degree of correlations of time-series patterns, thereby reflecting the interplay between order and disorder. Plotting the PE and the $C_{JS}[P]$ of a time-series in the CH-plane, one can distinguish between different dynamical behaviors: deterministic time-series are typically characterized by low entropy and high complexity, while purely random processes such as Brownian motion by high entropy and low complexity.

Furthermore, we employ the \textit{permutation Jensen-Shannon distance} (PJSD)~\cite{pJSdistance} to quantify the degree of similarity between each cryptocurrency time-series and various colored noises. This measure combines the concepts of Jensen-Shannon divergence ($D_{JS}$) and Permutation Entropy (PE) to provide a robust metric for comparing time-series.
The PJSD is calculated as
$
\sqrt{D_{JS}(P, Q)}
$
where $P$ and $Q$ are the ordinal probability distributions associated with the two time-series under analysis. In our case, $P$ represents the distribution of a cryptocurrency time-series, while $Q$ represents the distribution of a specific colored noise.
By computing the PJSD between each cryptocurrency and various colored noises, we can precisely quantify how closely the dynamics of cryptocurrency markets resemble different types of random processes. 

Finally, we evaluate the cryptocurrency time-series in the frequency domain through the \textit{Power Spectral Density} (PSD).
The PSD plot shows how the power of a signal is distributed over frequencies. It provides key insights into the periodicities, dominant frequencies, and scaling behaviors of the time-series. 
In particular, the power spectral density $S(f)$ of colored noises follows a power law distribution as a function of the frequency $f$: $S(f) \propto 1/f^{\alpha}$, where 
$\alpha$
is the power-law exponent characterising each colored noise.
For instance, a lower absolute value of $\alpha$ indicates 
that the time-series is close to white noise ($\alpha = 0$ being its characteristic exponent), while a higher absolute value of $\alpha$ indicates that the time-series reveals more structured patterns.

\subsection{Statistical, Machine and Deep Learning Models}

In this study, we employ a wide range of models for time-series forecasting, including statistical models, ML models, and DL models. Comparing the performance of different types of models enables us to gain insights into the underlying structure of the time-series, the complexity, the predictability, and the nature of the information present in the data. By comparing the performance of the different model types, we can infer whether the time-series exhibit identifiable patterns that can be learned and predicted, or if it is inherently noisy and lacks significant predictable signals. This approach enables a comprehensive assessment of cryptocurrency market dynamics and of the efficacy of various forecasting techniques in this financial domain.

The complete list of models used in our analysis is shown in Table~\ref{tab:models}.
The models and the back-testing procedure are implemented using the Darts~\citep{darts} Python library. 

\begin{table}[h!]
    \centering
    \caption{List of Models by Category}
    \begin{tabular}{ll}
        \toprule
        \textbf{Class of Models} & \textbf{Models} \\
        \midrule
        \multirow{5}{*}{Statistical} 
        & NaiveDrift~\citep{Hyndman2021}, NaiveSeasonal~\citep{Hyndman2021},\\
        & ARIMA~\citep{Hyndman2021},\\
        & Exponential Smoothing (ETS)~\citep{Hyndman2021} \\
        & Complex Exponential Smoothing (CES)~\citep{Svetunkov2017} \\
        & TBATS~\citep{DeLivera2011}, Prophet~\citep{Taylor2018}\\
        \midrule
        Machine Learning & RandomForest~\citep{Breiman2001}, XGBModel~\citep{Chen2016} \\
        \midrule
        Deep Learning & VanillaRNN~\citep{Elman1990}, LSTM~\citep{Hochreiter1997}, NBEATS~\citep{Oreshkin2019} \\
        \bottomrule
    \end{tabular}
    \label{tab:models}
\end{table}

The back-testing procedure trains a model $M$ with a specified forecast horizon $f_h$ and time window $t_w$. It begins by training $M$ on the initial $y_{train}$ dataset, constrained to the defined time window $t_w$. The model then generates a forecast spanning $f_h$ time steps. Following this, the procedure incrementally expands $y_{train}$ by incorporating one sample from $y_{target}$. This process repeats iteratively, continuously updating the training set and producing new forecasts. This rolling window approach enables a comprehensive evaluation of the model's predictive performance across various temporal segments of the data. More specifically, at time $t$, the forecast is obtained as:
\begin{displaymath}
    \tilde{y}(t) = M(y(t-fh), y(t-fh-1), \ldots, y(t-fh-t_w)).
\end{displaymath}

The forecasting metric used for comparing the models accuracy is the Mean Absolute Percentage Error (MAPE), defined as: 
\begin{displaymath}
    \text{MAPE}~(\tilde{y},y_{target})=\frac{1}{T} \sum_{t=1}^{T} \left|\frac{\tilde{y}(t) - y_{target}(t)}{y_{target}(t)}\right| \times 100
\end{displaymath}
where $\tilde{y}$ corresponds to the forecast of a given model, $t=1$ is the first forecast index and $T$ is the size of $y_{target}$. The MAPE metric is evaluated over the entire $y_{target}$ by using only the point predicted at the given forecast horizon $f_h$. This approach allows for a focused evaluation of the models' predictive accuracy at specific future time points, as shorter-forecast points do not influence the metric computation.

\section{Results}
\begin{figure}[ht!]
    \centering
    \begin{subfigure}{\linewidth}
        \includegraphics[width=\linewidth]{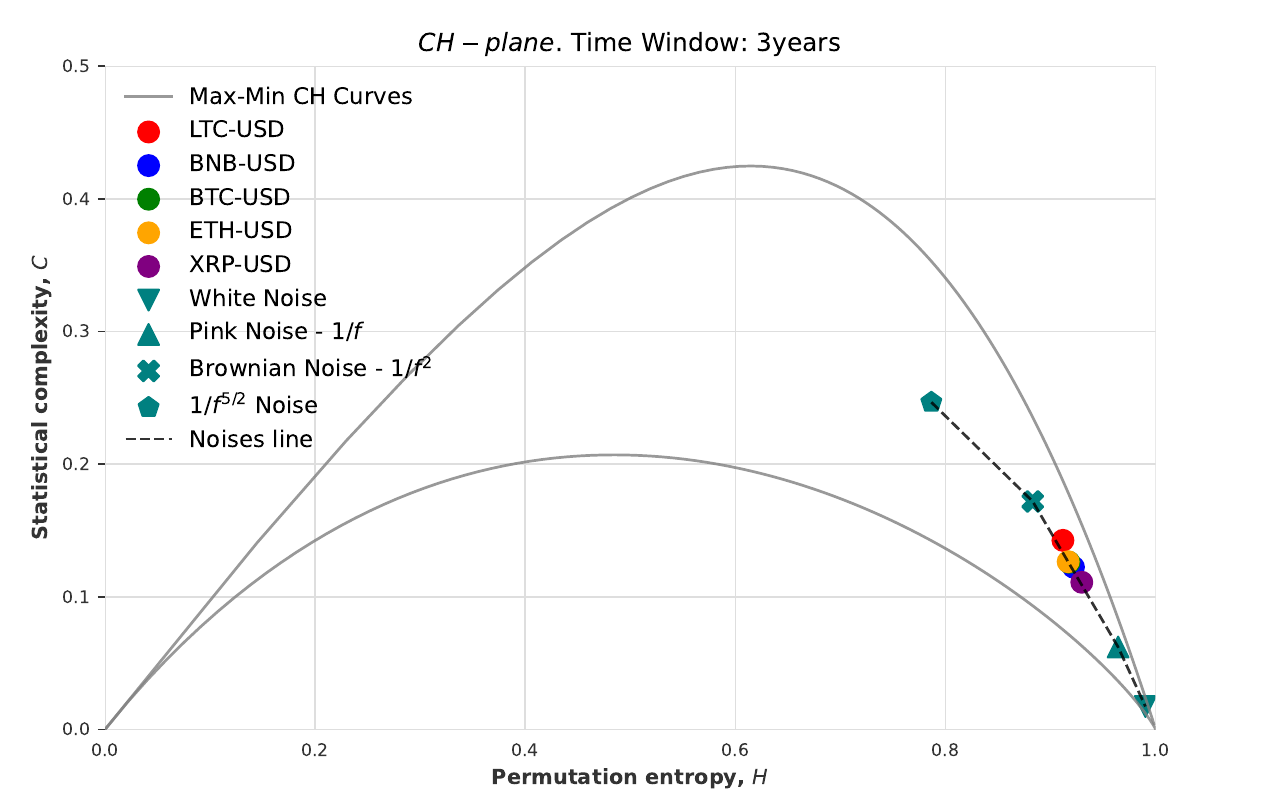}
        \caption{CH-plane: time window $t_w$ = 3 years}
        \label{fig:chplane3y}
    \end{subfigure}

    \begin{subfigure}{\linewidth}
        \includegraphics[width=\linewidth]{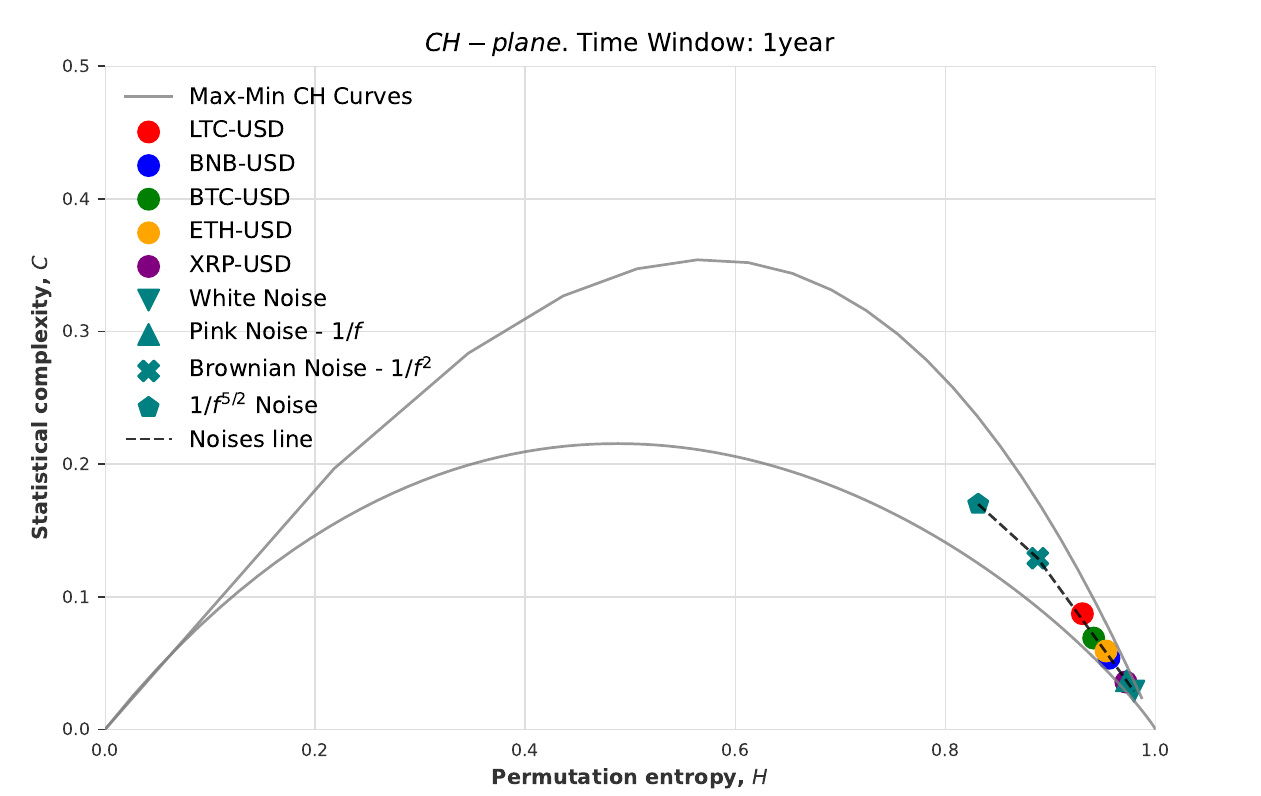}
        \caption{CH-plane: time window $t_w$ = 1 year}
        \label{fig:chplane1y}
    \end{subfigure}
    
    \begin{subfigure}{\linewidth}
        \includegraphics[width=\linewidth]{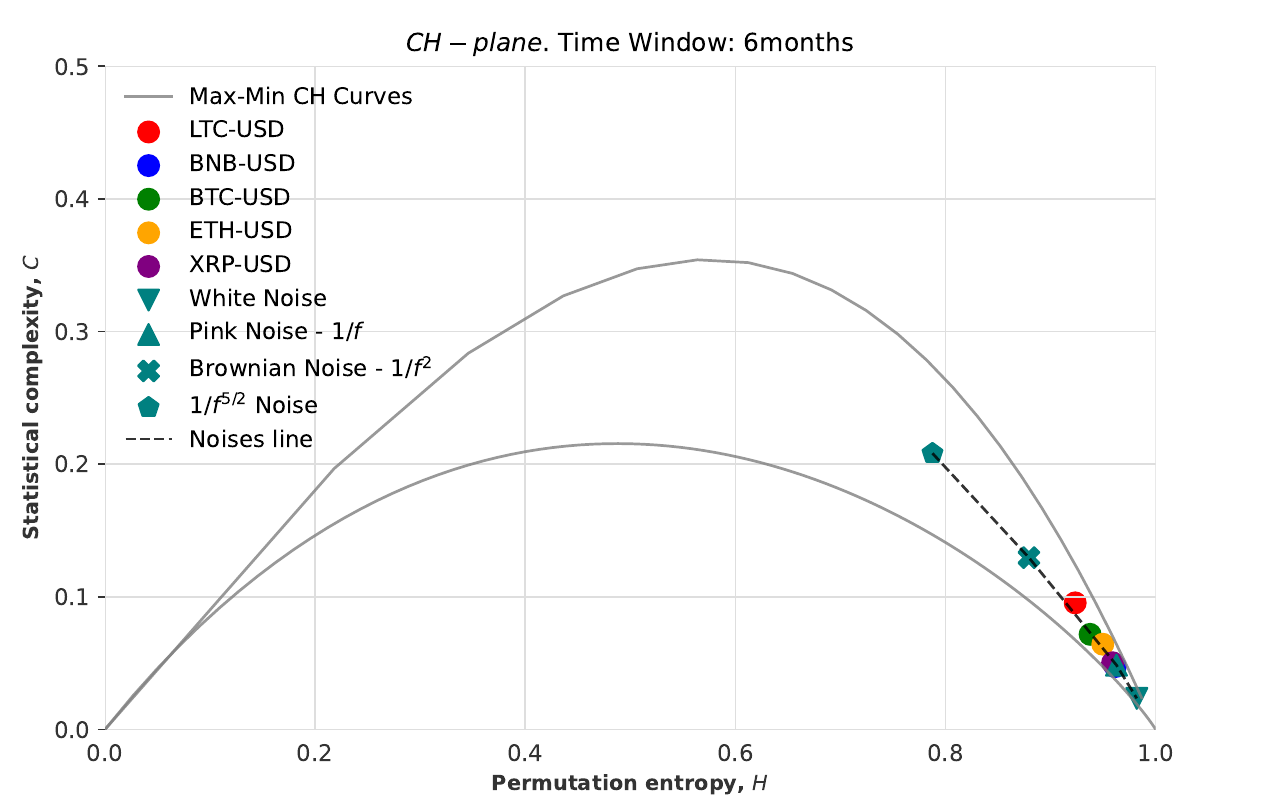}
        \caption{CH-plane: time window $t_w$ = 6 months}
        \label{fig:chplane6m}  
    \end{subfigure}
    
    \caption{Complexity Entropy causality plane (CH-plane) of cryptocurrencies time-series (LTC-USD, BNB-USD, BTC-USD, ETH-USD, XRP-USD) compared to different types of noises: white, $1/f$ (pink), $1/f^{2}$ (Brownian), and $1/f^{5/2}$, computed using different time windows settings. All the cryptocurrencies lie on the dashed line characterizing the colored noises. As the time window shortens, the cryptocurrency data increasingly align with the position of white noise.}
    \Description{}
    \label{fig:enter-label}
\end{figure}

\subsection{Complexity measures} 

The CH-plane plots depicted in Fig.~\ref{fig:chplane3y}, \ref{fig:chplane1y} and \ref{fig:chplane6m} show that all cryptocurrencies are situated in the lower right corner, indicating high permutation entropy and low statistical complexity. This positioning suggests that their statistical properties closely resemble those of different types of noise, i.e., colored noises such as white, $1/f$ (pink), $1/f^2$ (Brownian), and $1/f^{5/2}$ noises, which are represented in Fig.~\ref{fig:enter-label} along the dashed "noises line". Indeed, the cryptocurrency time-series lie precisely on the noises line. 

The high permutation entropy values imply a significant degree of randomness and unpredictability, while the relatively low statistical complexity values indicate minimal underlying structure. Moreover, as the time window shortens, i.e., from $t_w=3$ years to $t_w=6$ months, the statistical characteristics of the cryptocurrency time-series increasingly align with those of white noise. This behaviour suggests a growing level of randomness and unpredictability, and the lack of meaningful patterns in cryptocurrency time-series particularly over short-term periods.

To quantify these similarities more precisely, we computed the PJSD between each cryptocurrency and various types of colored noise over the full available time window 2020-07-03 to 2023-12-21. Results are shown in Table~\ref{tab:crypto-noise-distances} and provide a clear comparative analysis of which type of noise each cryptocurrency most closely resembles. The PJSD values indicate that, for most cryptocurrencies, the minimum distance is observed with respect to Brownian noise. However, an exception is noted with XRP-USD, which shows a higher similarity to pink noise.

\begin{table}[ht!]
    \caption{Permutation Jehnsenn-Shannon distances between Cryptocurrencies and colored noises over the full available time window 2020-07-03 to 2023-12-21. The minimum distance is observed with Brownian noise for most cryptocurrencies. XRP-USD represents an exception showing a higher similarity to pink noise.}

    \resizebox{\columnwidth}{!}{
        \begin{tabular}{cccccc}
\toprule
 & White & Pink - $1/f$ & Brownian - $1/f^2$ & $1/f^{5/2}$ & $1/f^3$ \\
\midrule
LTC-USD & 0.368 & 0.215 & \textbf{0.198} & 0.334 & 0.539 \\
BNB-USD & 0.358 & 0.212 & \textbf{0.209} & 0.353 & 0.552 \\
BTC-USD & 0.355 & 0.211 & \textbf{0.194} & 0.338 & 0.530 \\
ETH-USD & 0.356 & 0.202 & \textbf{0.198} & 0.339 & 0.536 \\
XRP-USD & 0.351 & \textbf{0.214} & 0.230 & 0.374 & 0.566 \\
\bottomrule
\end{tabular}

    }
    \label{tab:crypto-noise-distances}
\end{table}

\begin{figure}[ht]
    \includegraphics[width=\linewidth]{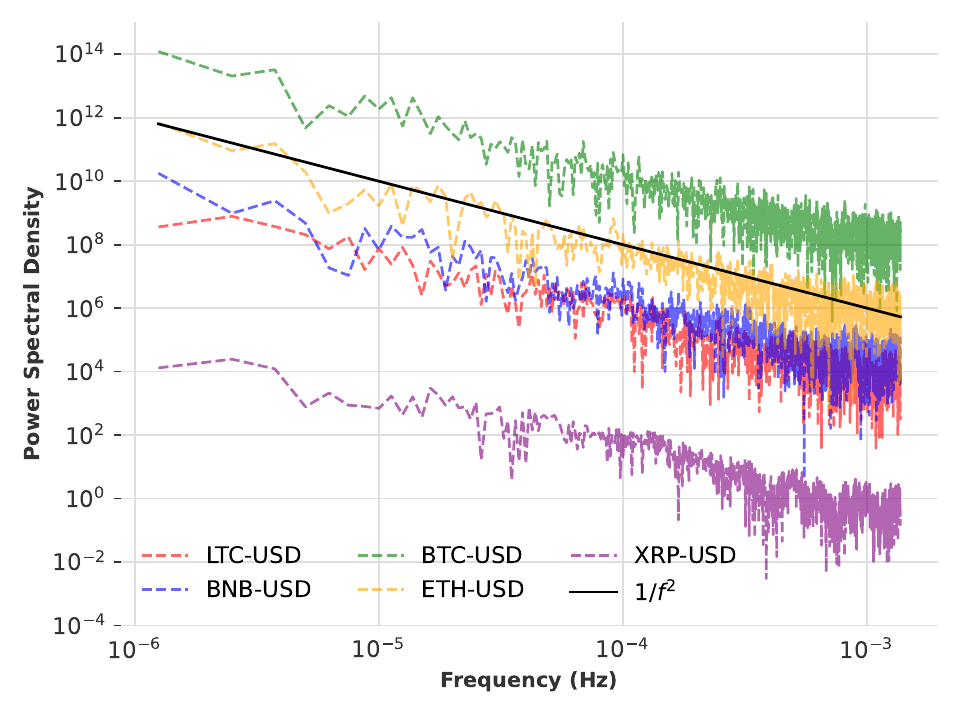}
    \caption{Power Spectral Density (PSD) plots of the 5 cryptocurrency univariate time-series (LTC-USD, BNB-USD, BTC-USD, ETH-USD, XRP-USD) computed using the whole available data time range (from 2020-07-03 to 2023-12-21). The PSD plots shows that cryptocurrency time-series follow a power-law distribution comparable to Brownian noise, whose exponent is indicated by the full black reference line.}
    \Description{}
    \label{fig:psd-plot} 
\end{figure}

The PSD curves shown in Fig.~\ref{fig:psd-plot} are computed by using the whole available data time range, i.e., from 2020-07-03 to 2023-12-21. The plot corroborates the similarity of the cryptocurrency time-series to $1/f^2$ (Brownian) noise whose exponent is represented by the full black reference line. 
In addition, the analysis was extended to shorter time windows, i.e. $t_w=3$ years, $t_w=1$ year, and $t_w=6$ months, as well. Interestingly, the spectral characteristics of the cryptocurrencies remain consistent, i.e., the PSD consistently follows the power law decay exponent of the Brownian motion, regardless of the time window.
The plot of the PSD for the other time windows is not reported here for brevity. 

\begin{table*}
    \centering
    \caption{
    MAPE results for statistical and ML/DL models across five cryptocurrencies (LTC-USD, BNB-USD, BTC-USD, ETH-USD, XRP-USD) for (a) 3-year, (b) 1-year, and (c) 6-month time windows, with forecast horizons of 1, 7, and 30 days. 
    A key finding is that statistical approaches consistently outperform ML and DL models, with the exception of NBEATS, which shows competitive but not consistently superior performance.
    Interestingly, sophisticated statistical models like AutoARIMA, AutoETS, AutoCES, and TBATS did not significantly outperform simpler Naive models.
    DL models, despite their reputation for modeling complex sequences, face challenges in cryptocurrency forecasting. As the time window shrinks, prediction errors for VanillaRNN and LSTM models increase across all forecast horizons. Similarly, traditional ML algorithms like RandomForest and XGBModel struggle to discern patterns amid the inherent noise and volatility.} 
    \label{tab:models-mape}
    \begin{subtable}{\textwidth}
        \caption{}
        \label{tab:3years}
        \resizebox{\linewidth}{!}{
            \begin{tabular}{llllllllllllllll}
\toprule
Time Window & \multicolumn{15}{c}{3 years} \\
Dataset & \multicolumn{3}{l}{LTC-USD} & \multicolumn{3}{l}{BNB-USD} & \multicolumn{3}{l}{BTC-USD} & \multicolumn{3}{l}{ETH-USD} & \multicolumn{3}{l}{XRP-USD}\\
Forecast Horizon & 1 & 7 & 30 & 1 & 7 & 30 & 1 & 7 & 30 & 1 & 7 & 30 & 1 & 7 & 30 \\
\midrule
NaiveDrift & 1.914 & 5.361 & 12.562 & 1.415 & 3.808 & 7.815 & 1.298 & 3.617 & \textbf{11.012} & 1.452 & 3.946 & 9.203 & 2.110 & 6.600 & 16.589 \\
NaiveSeasonal & \textbf{1.911} & 5.298 & 12.436 & 1.419 & 3.759 & 7.049 & 1.296 & 3.619 & 11.307 & 1.448 & 3.886 & \textbf{9.011} & 2.105 & 6.517 & 16.276 \\
AutoARIMA & 1.929 & 5.318 & 12.444 & 1.437 & 3.770 & 7.055 & 1.296 & 3.619 & 11.307 & \textbf{1.447} & 3.875 & 9.011 & 2.093 & \textbf{6.5} & \textbf{16.264} \\
AutoETS & 1.933 & 5.715 & 14.884 & 1.433 & 4.065 & 10.249 & \textbf{1.292} & \textbf{3.579} & 11.071 & 1.461 & 3.968 & 9.439 & 2.131 & 6.983 & 18.397 \\
AutoCES & 1.938 & \textbf{5.233} & 12.766 & \textbf{1.408} & \textbf{3.735} & \textbf{7.023} & 1.310 & 3.652 & 11.662 & 1.452 & \textbf{3.843} & 9.096 & \textbf{2.07} & 6.529 & 17.290 \\
TBATS & 1.912 & 5.294 & \textbf{12.429} & 1.410 & 3.832 & 7.685 & 1.293 & 3.618 & 11.310 & 1.447 & 3.880 & 9.011 & 2.094 & 6.506 & 16.267 \\
Prophet & 22.301 & 27.390 & 42.971 & 11.929 & 14.062 & 18.253 & 10.140 & 13.130 & 23.244 & 12.247 & 15.457 & 24.848 & 15.492 & 18.903 & 27.875 \\
RandomForest & 2.202 & 8.234 & 18.755 & 1.651 & 4.974 & 25.132 & 1.675 & 6.531 & 11.728 & 1.694 & 5.117 & 10.593 & 2.364 & 9.501 & 30.308 \\
XGBModel & 2.707 & 10.355 & 19.039 & 1.985 & 5.483 & 24.539 & 1.902 & 6.546 & 13.549 & 1.850 & 5.577 & 12.534 & 2.715 & 12.416 & 37.773 \\
VanillaRNN & 2.064 & 7.209 & 19.112 & 2.967 & 6.772 & 8.973 & 99.091 & 99.102 & 99.120 & 84.672 & 84.746 & 85.049 & 2.237 & 7.922 & 29.138 \\
LSTM & 2.195 & 7.461 & 19.425 & 2.564 & 7.541 & 9.226 & 99.765 & 99.789 & 99.794 & 93.190 & 93.211 & 93.349 & 2.333 & 8.359 & 31.029 \\
NBEATS & 2.717 & 7.730 & 16.043 & 2.023 & 5.680 & 23.357 & 1.875 & 4.801 & 15.742 & 2.006 & 5.512 & 12.853 & 3.393 & 13.202 & 32.761 \\
\bottomrule
\end{tabular}
          
        }
    \end{subtable}
    \par\bigskip        
    \begin{subtable}{\textwidth}
        \caption{}
        \label{tab:1year}
        \resizebox{\linewidth}{!}{
            \begin{tabular}{llllllllllllllll}
\toprule
Time Window & \multicolumn{15}{c}{1 year} \\
Dataset & \multicolumn{3}{l}{LTC-USD} & \multicolumn{3}{l}{BNB-USD} & \multicolumn{3}{l}{BTC-USD} & \multicolumn{3}{l}{ETH-USD} & \multicolumn{3}{l}{XRP-USD} \\
Forecast Horizon & 1 & 7 & 30 & 1 & 7 & 30 & 1 & 7 & 30 & 1 & 7 & 30 & 1 & 7 & 30 \\
\midrule
NaiveDrift & 1.919 & 5.458 & 12.995 & 1.420 & 3.779 & 7.277 & 1.299 & 3.627 & 11.077 & 1.455 & 3.967 & 9.390 & 2.117 & 6.703 & 17.179 \\
NaiveSeasonal & \textbf{1.911} & \textbf{5.298} & 12.436 & 1.419 & 3.759 & 7.049 & 1.296 & 3.619 & 11.307 & 1.448 & 3.886 & \textbf{9.011} & 2.105 & \textbf{6.517} & 16.276 \\
AutoARIMA & 1.911 & 5.298 & 12.436 & 1.404 & 3.741 & 7.035 & 1.296 & 3.619 & 11.307 & 1.449 & 3.877 & 9.011 & 2.103 & 6.527 & 16.278 \\
AutoETS & 1.911 & 5.298 & 12.436 & \textbf{1.402} & \textbf{3.735} & \textbf{7.032} & 1.296 & 3.619 & 11.307 & 1.447 & \textbf{3.875} & 9.012 & 2.079 & 6.555 & 16.243 \\
AutoCES & 1.945 & 5.335 & 13.348 & 1.412 & 3.750 & 7.104 & 1.315 & 3.612 & 11.230 & 1.457 & 3.881 & 9.150 & 2.104 & 6.743 & 17.514 \\
TBATS & 1.914 & 5.306 & \textbf{12.271} & 1.429 & 4.067 & 9.189 & \textbf{1.293} & \textbf{3.595} & \textbf{10.605} & \textbf{1.44} & 3.903 & 9.226 & \textbf{2.071} & 6.606 & 17.082 \\
Prophet & 11.709 & 14.065 & 22.853 & 6.931 & 8.420 & 15.934 & 7.437 & 9.022 & 15.100 & 7.054 & 8.407 & 13.501 & 12.627 & 14.862 & 21.915 \\
RandomForest & 2.080 & 6.878 & 19.199 & 1.685 & 4.866 & 17.374 & 1.693 & 5.955 & 12.300 & 1.773 & 4.597 & 10.449 & 2.699 & 9.355 & 20.851 \\
XGBModel & 2.499 & 8.870 & 20.802 & 1.875 & 5.475 & 18.333 & 2.029 & 6.914 & 11.978 & 1.963 & 5.012 & 11.227 & 3.082 & 10.037 & 24.186 \\
VanillaRNN & 8.156 & 11.960 & 12.618 & 56.079 & 56.673 & 58.750 & 99.658 & 99.674 & 99.691 & 94.179 & 94.339 & 94.640 & 2.219 & 6.955 & \textbf{16.117} \\
LSTM & 8.217 & 10.302 & 13.364 & 77.060 & 78.361 & 79.712 & 99.912 & 99.929 & 99.936 & 98.403 & 98.613 & 98.772 & 2.127 & 6.824 & 17.304 \\
NBEATS & 2.573 & 7.270 & 16.216 & 2.015 & 4.593 & 9.293 & 1.744 & 4.741 & 13.307 & 1.835 & 4.727 & 11.594 & 3.283 & 9.125 & 22.532 \\
\bottomrule
\end{tabular}
        
        }
    \end{subtable}
    \par\bigskip       
    \begin{subtable}{\textwidth}
        \caption{}
        \label{tab:6months}
        \resizebox{\linewidth}{!}{
            \begin{tabular}{llllllllllllllll}
\toprule
Time Window & \multicolumn{15}{c}{6 months} \\
Dataset & \multicolumn{3}{l}{LTC-USD} & \multicolumn{3}{l}{BNB-USD} & \multicolumn{3}{l}{BTC-USD} & \multicolumn{3}{l}{ETH-USD} & \multicolumn{3}{l}{XRP-USD} \\
Forecast Horizon & 1 & 7 & 30 & 1 & 7 & 30 & 1 & 7 & 30 & 1 & 7 & 30 & 1 & 7 & 30 \\
\midrule
NaiveDrift & 1.922 & 5.454 & 13.676 & 1.425 & 3.803 & 7.433 & 1.315 & 3.787 & 11.247 & 1.463 & 4.062 & 9.966 & 2.130 & 6.881 & 18.209 \\
NaiveSeasonal & 1.911 & 5.298 & 12.436 & 1.419 & 3.759 & 7.049 & \textbf{1.296} & 3.619 & 11.307 & \textbf{1.448} & 3.886 & 9.011 & 2.105 & \textbf{6.517} & 16.276 \\
AutoARIMA & \textbf{1.907} & \textbf{5.205} & \textbf{11.907} & 1.401 & 3.759 & 8.460 & 1.296 & 3.619 & 11.307 & 1.456 & 3.871 & 9.014 & 2.115 & 6.565 & 16.261 \\
AutoETS & 1.911 & 5.298 & 12.435 & \textbf{1.4} & \textbf{3.742} & \textbf{7.043} & 1.298 & \textbf{3.617} & 11.157 & 1.449 & \textbf{3.869} & \textbf{9.007} & \textbf{2.073} & 6.578 & 16.230 \\
AutoCES & 1.948 & 5.359 & 13.823 & 1.414 & 3.763 & 7.291 & 1.333 & 3.765 & 11.306 & 1.472 & 3.992 & 9.788 & 2.138 & 6.990 & 18.763 \\
TBATS & 2.005 & 6.319 & 14.243 & 1.471 & 3.809 & 8.267 & 1.315 & 3.784 & \textbf{10.292} & 1.487 & 4.378 & 10.124 & 2.088 & 6.631 & 16.286 \\
Prophet & 8.903 & 11.634 & 23.436 & 5.173 & 6.805 & 12.722 & 5.532 & 7.451 & 16.171 & 4.631 & 6.163 & 12.856 & 10.751 & 13.959 & 25.909 \\
RandomForest & 2.417 & 8.036 & 18.194 & 1.652 & 4.149 & 14.504 & 1.655 & 6.077 & 12.362 & 1.826 & 4.498 & 9.907 & 2.728 & 9.237 & 20.008 \\
XGBModel & 2.689 & 9.011 & 20.523 & 1.873 & 5.081 & 15.371 & 1.803 & 6.350 & 11.773 & 2.082 & 4.771 & 10.499 & 3.947 & 9.819 & 24.050 \\
VanillaRNN & 25.377 & 26.717 & 31.597 & 73.680 & 74.399 & 76.592 & 99.801 & 99.820 & 99.838 & 96.590 & 96.848 & 97.161 & 2.279 & 6.761 & \textbf{15.244} \\
LSTM & 27.644 & 28.832 & 33.609 & 86.521 & 87.064 & 88.554 & 99.949 & 99.966 & 99.975 & 99.096 & 99.380 & 99.550 & 2.408 & 6.637 & 18.355 \\
NBEATS & 2.736 & 6.877 & 15.221 & 1.782 & 4.293 & 9.635 & 1.847 & 5.061 & 12.649 & 1.984 & 4.599 & 10.543 & 3.854 & 10.144 & 21.714 \\
\bottomrule
\end{tabular}

        }
    \end{subtable}     
\end{table*}

\begin{table}
    \centering
    \caption{
    Aggregated view of mean MAPE ± standard deviation across all cryptocurrencies and time windows corresponding to tables (a), (b) and (c) in Table~\ref{tab:models-mape}, for statistical, machine learning (ML), and deep learning (DL) models. This view illustrates that while statistical approaches generally outperform ML and DL methods, the complex statistical techniques (AutoARIMA, AutoETS, AutoCES, and TBATS) show no statistically significant improvement over Naive models, as evidenced by the overlapping standard deviations.}
    \label{tab:mapestatistics}
    \resizebox{0.98\linewidth}{!}{
        \begin{tabular}{llll}
\toprule
Forecast Horizon & 1 & 7 & 30 \\
\midrule
NaiveDrift & 1.644 ± 0.328 & 4.723 ± 1.227 & 11.709 ± 3.496 \\
NaiveSeasonal & 1.636 ± 0.325 & 4.616 ± 1.166 & \textbf{11.216 ± 3.251} \\
AutoARIMA & 1.636 ± 0.326 & \textbf{4.611 ± 1.17} & 11.273 ± 3.128 \\
AutoETS & \textbf{1.634 ± 0.324} & 4.7 ± 1.245 & 11.729 ± 3.415 \\
AutoCES & 1.648 ± 0.326 & 4.679 ± 1.244 & 11.81 ± 3.807 \\
TBATS & 1.645 ± 0.321 & 4.768 ± 1.21 & 11.619 ± 3.064 \\
Prophet & 10.19 ± 4.614 & 12.649 ± 5.532 & 21.173 ± 7.865 \\
RandomForest & 1.986 ± 0.397 & 6.534 ± 1.903 & 16.778 ± 5.903 \\
XGBModel & 2.333 ± 0.606 & 7.448 ± 2.423 & 18.412 ± 7.345 \\
VanillaRNN & 49.937 ± 43.905 & 51.993 ± 41.825 & 56.243 ± 37.961 \\
LSTM & 53.426 ± 45.967 & 55.485 ± 43.899 & 60.13 ± 39.327 \\
NBEATS & 2.378 ± 0.679 & 6.557 ± 2.567 & 16.231 ± 6.4 \\
\bottomrule
\end{tabular}

    }
\end{table}

\subsection{Model performances}
By examining Tables~\ref{tab:3years}, \ref{tab:1year} and \ref{tab:6months}, several observations can be made regarding the prediction accuracy of the considered models across cryptocurrencies, time windows~$t_w$, and forecast horizons~$f_h$. Notably, as shown in Table~\ref{tab:models-mape}, one of our core results is that statistical approaches consistently outperform their ML and DL counterparts.

Focusing on the statistical approaches, models like AutoARIMA, AutoETS, AutoCES, and TBATS did not significantly over-perform the much simpler Naive models. In fact, despite the greater sophistication and theoretical guarantees of such models, the fitting procedures of these advanced models resulted in average performances that are not statistically distinguishable from those achieved by the Naive models. The aggregated view in Table~\ref{tab:mapestatistics}, where the mean MAPE over all cryptocurrencies and all $t_w$ is reported, together with the corresponding standard deviations, substantiates this result.

In contrast, the Prophet model resulted in much worse performances compared to the other statistical methods, and in some cases, even compared to ML and DL models. This discrepancy can be attributed to the specific use-cases Prophet is designed to address, characterized by strong seasonal effects and holidays. Hence, the high volatility and the absence of regular seasonal patterns in the cryptocurrency markets present a challenging environment for Prophet’s underlying assumptions and mechanisms. 

DL models, renowned for their ability to model complex sequences and capture long-term dependencies, are often seen as holding the potential of superior performance in time series forecasting. However, their application to univariate cryptocurrency time series reveals inherent challenges. These models require substantial amounts of data to be trained effectively and are prone to overfitting, especially when compared to the noise-dominated nature of cryptocurrency data. These limitations become evident when examining the MAPE across decreasing time windows, as can be seen in Table~\ref{tab:models-mape}. The results clearly indicate that as the time window~$t_w$ shrinks, the prediction error of VanillaRNN and LSTM models increases, irrespective of the forecast horizon $f_h$. The NBEATS model stands out as a notable exception. However, while it delivers competitive prediction accuracy, its performance does not statistically match the top statistical models. This unexpected result underscores a critical insight: even advanced DL architectures may struggle in domains characterized by high volatility and noise. 

Akin to deep learning models, traditional ML algorithms like RandomForest and XGBModel - often acclaimed for their capacity to capture intricate non-linear relationships and maintain robust generalization - underperform in univariate cryptocurrency forecasting. These models also encounter difficulty in discerning significant patterns amid the stochastic fluctuations and inherent noise of cryptocurrency markets.

\section{Discussion and Conclusion}

Our study reveals that univariate forecasting of cryptocurrencies is essentially comparable to pure noise forecasting. Simpler statistical models are consistently comparable or outperform more complex ML and DL models across various forecast horizons and time windows in an extended time range from 2020-07-03 to 2023-12-21, for five prominent cryptocurrencies, i.e. LTC-USD, BNB-USD, BTC-
USD, ETH-USD and XRP-USD. 
Complexity analysis using the CH-plane and the Power Spectral Density (PSD) highlights the noisy nature of cryptocurrency time-series, revealing high entropy, low complexity, and PSD power law exponents comparable with those of Brownian motion. These findings collectively demonstrate the inherent stochastic nature of cryptocurrencies and the varying degrees of noise-like behavior they exhibit over different time scales.
These insights challenge the presence of predictable patterns in cryptocurrency markets and suggest that their apparent complexity may be largely attributed to noise. The resemblance to Brownian motion implies that forecasting future prices based solely on historical data may be unfeasible. 
Likewise, the similarity to white noise over shorter periods points to increased randomness and potential challenges even in short-term forecasting.

This study challenges the conventional wisdom that increased model complexity guarantees better performance. The inherent unpredictability and rapid evolution of cryptocurrency markets pose significant hurdles for deep learning and machine learning models. These models, often acclaimed for their sophisticated designs, may not consistently deliver superior performance across all contexts. This aligns with the findings in~\citep{LahmiriChaos}, which observed that simple forecasting methods can outperform more complex ones in cryptocurrency markets.
Our study invites researchers and practitioners to reconsider their approach to model selection, emphasizing the value of simple models for what concerns the ability to handle noise and volatility. 

However, it's important to note that incorporating additional covariates can significantly improve the forecasting accuracy of cryptocurrency models. These covariates can be categorized into past covariates, such as technical indicators and correlated time-series data, and future covariates, including scheduled events and macroeconomic forecasts. Past covariates may help mitigate noise and complexity in the models, while future covariates can assist in anticipating external influences on cryptocurrency markets.
While past covariates are generally easier to obtain and incorporate, future covariates present some challenges. They are often difficult to retrieve directly, and when forecasted rather than known with certainty, they can introduce additional uncertainty into the model. This observation is consistent with the findings of~\citep{extraregr}, which demonstrated improved forecasting performance by integrating additional market indicators and sentiment analysis. Similarly,~\citep{catania2019forecasting} showed that incorporating exogenous variables can enhance cryptocurrency volatility forecasting.

In conclusion, our study highlights the importance of balancing model complexity with the inherent noise and unpredictability in cryptocurrency markets. While more sophisticated models may offer potential benefits, the effectiveness of simpler models should not be underestimated. Future research should focus on identifying the most relevant and impactful covariates for cryptocurrency forecasting, as well as developing methods to effectively incorporate future covariates without introducing excessive uncertainty.

\begin{acks}
This paper is supported by the PNRR-PE-AI FAIR project funded by the NextGeneration EU program and by Dhiria srl.
\end{acks}

\bibliographystyle{ACM-Reference-Format}
\bibliography{mybibfile}

\end{document}